\documentclass[sn-mathphys-num]{sn-jnl}

\usepackage{graphicx}%
\usepackage{multirow}%
\usepackage{amsmath,amssymb,amsfonts}%
\usepackage{amsthm}%
\usepackage{mathrsfs}%
\usepackage[title]{appendix}%
\usepackage{xcolor}%
\usepackage{textcomp}%
\usepackage{manyfoot}%
\usepackage{booktabs}%
\usepackage{algorithm}%
\usepackage{algorithmicx}%
\usepackage{algpseudocode}%
\usepackage{listings}%

\usepackage{lineno}
\usepackage{lscape}
\usepackage{adjustbox}
\usepackage{rotating} 
\usepackage{tabularx}
\usepackage{xurl}
\usepackage{makecell}

\theoremstyle{thmstyleone}%
\theoremstyle{thmstyletwo}%

\theoremstyle{thmstylethree}%

\begin{document}                                                                                                      

\title[Article Title]{Quantifying the Dynamics of Harm Caused by Retracted Research}


\author{
	Yunyou~Huang$^{1,4,8,10}$\
	Jiahui~Zhao$^{1,4}$, Dandan~Cui$^{2}$, Zhengxin~Yang$^{2,9}$,
        Bingjie~Xia$^{6}$,
        Qi~Liang$^{1,4}$, 
        Wenjing Liu$^{7}$, 
        Li Ma$^{7}$,
        Suqin~Tang$^{1}$,Tianyong~Hao$^{3,\ast}$,
        Zhifei~Zhang$^{5,\ast}$,
        Wanling~Gao$^{2,9,10,\ast}$,
	Jianfeng~Zhan$^{2,9,10,\ast}$
}

\affil[1]{\orgdiv{Key Lab of Education Blockchain and Intelligent Technology, Ministry of Education}, 
\orgname{Guangxi Normal University}, \orgaddress{\city{Guilin}, \postcode{541004}, \country{China}}}

\affil[2]{\orgdiv{Institute of Computing Technology}, \orgname{Chinese Academy of Sciences}, \orgaddress{\city{Beijing}, \postcode{100190}, \country{China}}}

\affil[3]{\orgdiv{School of Computer Science}, \orgname{South China Normal University}, \orgaddress{\city{Guangzhou}, \postcode{510631}, \country{China}}}

\affil[4]{\orgdiv{Guangxi Key Lab of Multi-Source Information Mining and Security}, \orgname{Guangxi Normal University}, \orgaddress{\city{Guilin}, \postcode{541004}, \country{China}}}

\affil[5]{\orgdiv{Department of Physiology and Pathophysiology}, \orgname{Capital Medical University}, \city{Beijing}, \postcode{100069}, \country{China}}

\affil[6]{\orgdiv{The second Affiliated hospital of Guilin medical university}, \city{Guilin}, \postcode{541004}, \country{China}}

\affil[7]{\orgdiv{Guilin Medical University}, \city{Guilin}, \postcode{541100}, \country{China}}

\affil[8]{\orgdiv{XuanJi Technology Co., Ltd.}, \city{Guilin}, \postcode{541012}, \country{China}}

\affil[9]{\orgdiv{University of Chinese Academy of Sciences}, \city{Beijing}, \postcode{100086}, \country{China}}

\affil[10]{\orgdiv{The International Open Benchmark Council}, \city{Delaware}, \postcode{19801}, \country{USA}}

\affil[*]{Corresponding authors: Jianfeng Zhan (zhanjianfeng@ict.ac.cn) and Wanling Gao (gaowanling@ict.ac.cn) and Zhifei Zhang (zhifeiz@ccmu.edu.cn) and Tianyong Hao (haoty@m.scnu.edu.cn)}








\abstract{Despite enormous efforts devoted to understanding the characteristics and impacts of retracted papers, little is known about the mechanisms underlying the dynamics of their harm and the dynamics of its propagation. Here, we propose a citation-based framework to quantify the harm caused by retracted papers, aiming to uncover why their harm persists and spreads so widely. We uncover an ``attention escape'' mechanism, wherein retracted papers postpone significant harm, more prominently affect indirectly citing papers, and inflict greater harm on citations in journals with an impact factor less than $10$. This mechanism allows retracted papers to inflict harm outside the attention of authors and publishers, thereby evading their intervention. This study deepens understanding of the harm caused by retracted papers, emphasizes the need to activate and enhance the attention of authors and publishers, and offers new insights and a foundation for strategies to mitigate their harm and prevent its spread.}

\maketitle

\section{Introduction}\label{sec1}

As of January 31, 2025, more than 60,000 papers have been publicly retracted, and this staggering number may only represent the tip of the iceberg~\cite{retraction_database1228,van2023more}. These papers have even dealt a fatal blow to entire research fields, as exemplified by the devastating cases of fraud in stem cell research~\cite{RetractionWatch2014} and Alzheimer’s research~\cite{piller2022blots}. Worse still, in recent years, the number of retracted papers has remained remarkably high each year, making the challenges we face increasingly severe~\cite{brainard2018massive,else2024biomedical,retraction_database1228}. 


Research on retracted papers has made significant progress, but most efforts have focused on their bibliometric and textual characteristics~\cite{RetractionWatch,PubPeer,ScienceIntegrityDigest,maurya2024reasons,sharma2023systematic,sevryugina2023analysis,peng2022dynamics,candal2022retracted,ferraro2022characteristics,heibi2021qualitative,yang2024retraction,heibi2022quantitative,wadhwa2021temporal,lin2024trend}, whereas relatively little attention has been paid to the harm these papers cause and the mechanisms by which such harm spreads~\cite{fanelli2022difference,stern2014financial}. Currently, researchers widely believe that the persistent harm of retracted papers stems from the difficulty authors, reviewers, and journals face in identifying and tracking their citations~\cite{bucci2019zombie,cabanac2024chain,swogger2024interactive}. To address this issue, tools such as Feet of Clay Detector~\cite{nature_3ysnj8f}, Annulled Detector~\cite{dbrech}, and RetractoBot~\cite{feakins2019retractobot} have been developed to flag citations of retracted papers and alert authors to handle such references with caution. However, recent studies have suggested that these measures have been less effective than anticipated. One study found that the act of flagging citations of retracted papers provoked resistance from some authors~\cite{van2024exclusive}. Another study flagged 88 articles citing retracted papers and sent notification emails to the authors, but only 44 responded, and just 11 implemented corrective actions~\cite{avenell2024randomized}. These findings indicated that current understanding of the harm caused by retracted papers remains significantly limited and biased. To mitigate the long-term impact of retracted literature, more in-depth research and fresh perspectives are urgently needed to drive meaningful progress in this field.

Citations are a key and typical form through which the harm of retracted papers arises and spreads. At the same time, citations, like publications, are regarded as the “currency” of academia, forming the foundation of scholarly influence~\cite{cabanac2024chain}. They bolster the reputation of authors and institutions, drive career advancement, secure funding opportunities, and contribute to journal rankings and impact metrics~\cite{sinatra2016quantifying,petersen2014reputation,van2014publication,wright2017faculty}. Recognizing this dual role, we propose a citation-based quantitative framework to deepen understanding of the harm caused by retracted papers and uncover the mechanisms driving it (see “Methods” section for details). To this end, we analyzed a refined dataset of 210,309,367 publications across 23 disciplines, extracted from the Semantic Scholar dataset following data cleaning and preprocessing.(\textit{SSD}, see “Methods” section)~\cite{kinney2023semantic}. By integrating data from the Retraction Watch dataset (\textit{RWD})~\cite{RetractionWatch}, which, after data cleaning and preprocessing, cataloged 39,449 retracted papers, we constructed an extensive citation network of retracted publications. This network enabled us to systematically trace the pathways through which harm originates and propagates, uncovering patterns that link retracted papers to their broader impact on the research community.

\begin{figure} 
	\centering
	\includegraphics[width=1\textwidth]{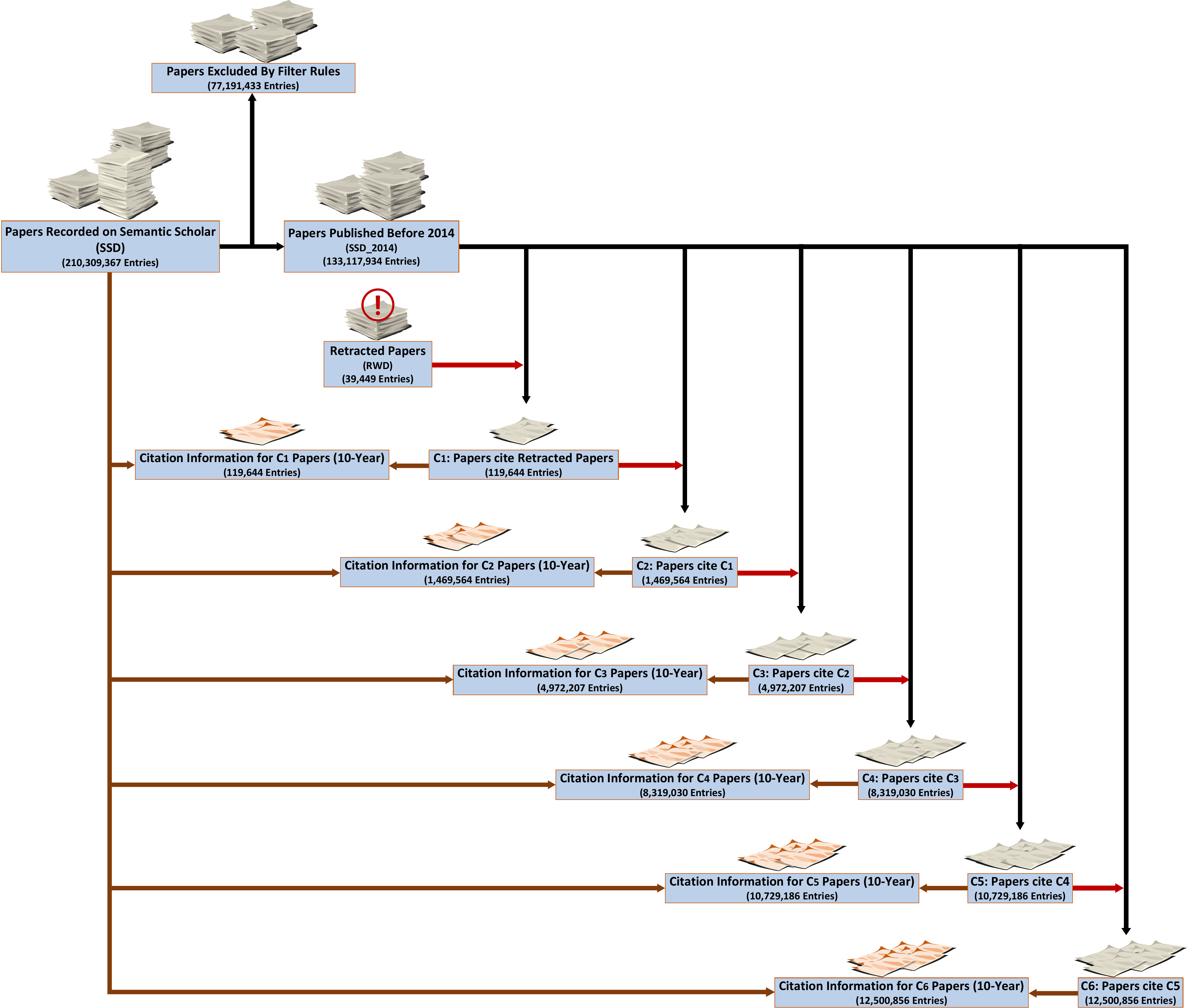}
	\caption{\textbf{Study Flowchart.} Description of the process of constructing the dataset of papers that directly cite retracted papers and indirectly cite retracted papers}
	\label{fig:paper_analysis}
\end{figure}

As illustrated in Figure~\ref{fig:paper_analysis}, to investigate the long-term impact of retracted papers, we focused on papers published before 2014 within the \textit{SSD}, creating a subset dataset named \textit{SSD\_{2014}} to track 10 years of citation dynamics. From the \textit{SSD\_{2014}} dataset, we identified papers directly citing those in the \textit{RWD} to form the citation distance 1 set (\(C_1\)). Subsequently, we iteratively extracted papers citing \(C_1\) to form the citation distance 2, 3, 4, 5, and 6 sets, denoted as \(C_2\), \(C_3\), \(C_4\), \(C_5\), and \(C_6\), respectively. For each of these sets, we calculated annual citation counts over 10 years and determined the expected citation counts based on contemporaneous papers published in the same journals and fields. The shortfall between observed and expected citations was defined as the harm caused by retracted papers, providing a quantitative measure of their harm and the dissemination of the harm. Details of the methodology are provided in the “Methods” section. Our research uncovered an “attention escape" mechanism by which retracted papers cause long-term, widespread, and significant harm to the scientific community. This mechanism operates through three key strategies: (1) Papers citing retracted studies experience insignificant harm during the early years when they receive the most attention, but this harm becomes increasingly significant over time. (2) Papers that indirectly cite retracted papers suffer more significant harm compared to those that directly cite them, and the severity of this harm increases with greater citation distance. (3) Papers citing retracted papers that are published in journals with an impact factor less than 10 experience more significant harm compared to those published in higher-impact, more widely recognized journals. This mechanism provides a new understanding of the harm caused by retracted papers, prompting us to shift our focus toward reactivating and enhancing the attention of authors, journals, and other stakeholders~\cite{zhan2024evaluatology}, thereby offering a fresh perspective for exploring new measures to mitigate the harm of retracted papers.

\section{Results}\label{sec2}

\begin{figure} 
	\centering
	\includegraphics[width=0.82\textwidth]{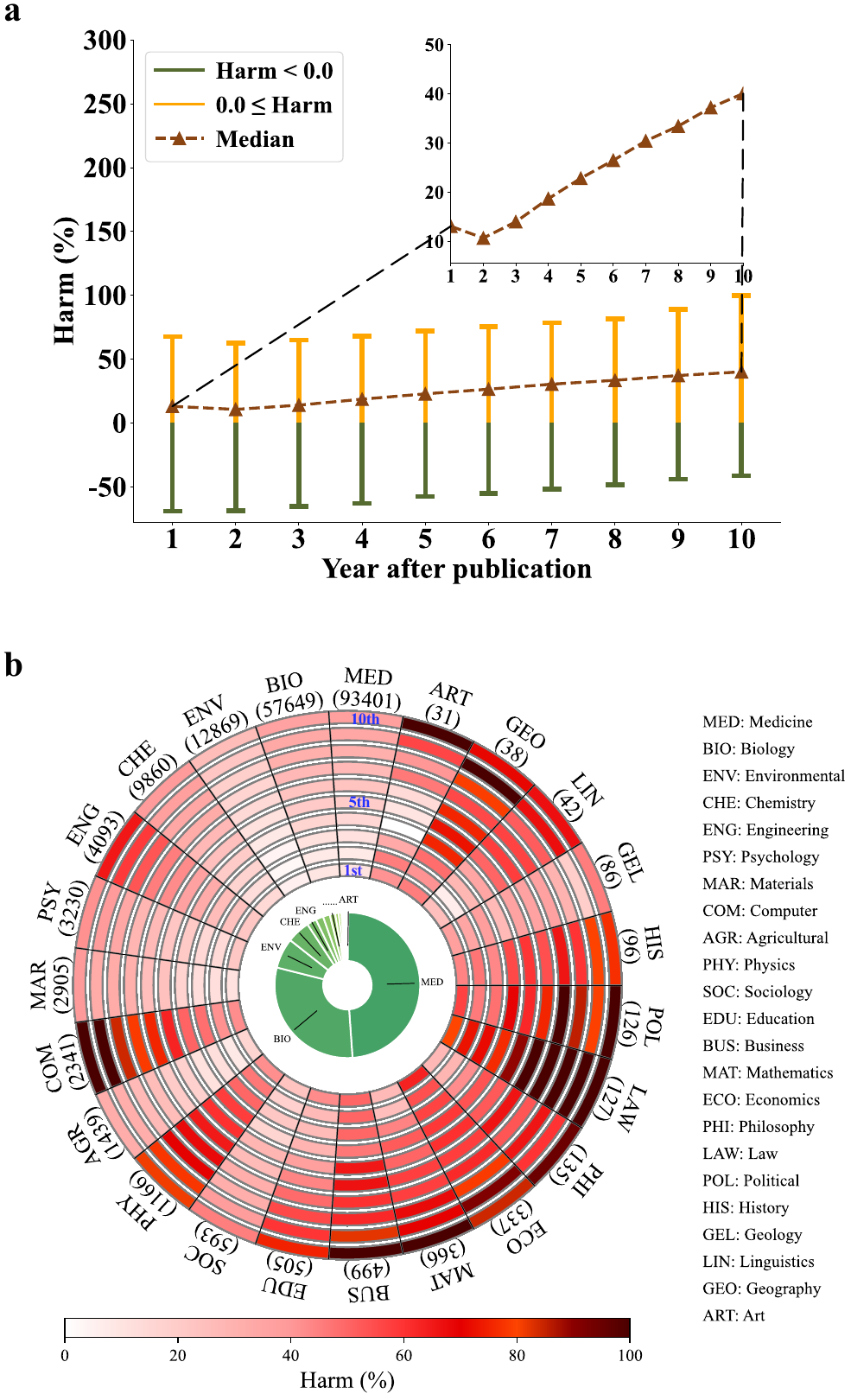} 

	\caption{\textbf{The variation in harm experienced by papers citing retracted articles over time.} \textbf{(a)} The overall variation in the median harm experienced by papers citing retracted articles over time. \textbf{(b)} The variation in the median harm experienced by papers citing retracted articles over time across different fields, with layers ranging from the first year post-publication (innermost) to the tenth year post-publication (outermost).}
	\label{fig:fig2_final} 
\end{figure}


\subsection{Temporal dynamics of harm arising from retracted papers}


To investigate how the harm of retracted papers evolves over time, we analyzed the annual citation impact within 10 years of publication for papers that cited these retracted works. Notably, the “10 years after publication" refers to the period starting from the second year to the eleventh year after the citing papers were published. The first year was excluded from the analysis because publication dates can vary widely throughout the year (e.g., January or December), which may result in significant biases in citation data for the first year. As shown in Figure~\ref{fig:fig2_final}\textbf{(a)}, the harm to papers directly citing retracted papers during the first four years after publication was not significant, with the highest median (interquartile range) harm observed in the fourth year, at 18.67\% (-62.92\% to 68.09\%). Starting in the fifth year, the harm to papers directly citing retracted papers exceeded 20\%, with a median (interquartile range) of 22.84\% (-57.55\% to 71.98\%). The lowest median (interquartile range) damage to the paper directly citing retracted papers was recorded in the second year at 10.68\% (-68.71\% to 62.46\%). Beyond the second year, the harm to the paper directly citing retracted papers increased annually, although the rate of increase slowed over time, reaching a median (interquartile range) of 40.08\% (-41.39\% to 100\%) by the tenth year. These findings suggest that the harm caused by retracted papers to citing papers accumulates over time, making long-term impacts unavoidable. We further analyzed the field-specific effects of retracted papers using Semantic Scholar’s discipline classifications. As shown in Figure~\ref{fig:fig2_final}\textbf{(b)}, disciplines with more than 300 citing articles (representing 99.65\% of the dataset) exhibited trends of harm to papers that directly cite retracted papers similar to the overall pattern. However, certain fields—such as computer science, physics, business, and economics (representing 2.26\% of the dataset)—showed significant harm to papers that directly cite retracted papers earlier in the citing papers’ lifespans. It should be noted that a single article may belong to multiple fields, so the total percentages exceed 100\%. Detailed statistics for each discipline are provided in Extended Data Table~\ref{tab:tableS1}.

Notably, in exploring the reasons behind the lack of attention to the harmful effects of problematic papers, we identified two phenomena potentially linked to the insufficient focus on the damaging mechanisms we uncovered, based on the interests most relevant to stakeholders. Two primary actors contributed to the dissemination of damage: authors (who cited retracted papers) and journals (which published papers citing retracted works). In our investigation of authors, we focused on universities as key research institutions. As shown in Table~\ref{tab:tableS6}, among the top 30 universities in the world, we found that 28 institutions placed significant emphasis on recent research outputs (typically within the past 3 to 5 years) when considering faculty promotions~\cite{world_university_rankings_2024}. In our investigation of journals, we found that most journals prioritized impact factor as a key metric of interest, with impact factor calculations typically based on papers published within the past 3 years~\cite{webofscience}. These phenomena suggested that papers from earlier years received far less attention from stakeholders, which could have been a key reason why the potential harm of problematic papers was overlooked.


Our study reveals that although citing retracted papers causes persistent, significant, and widespread harm to stakeholders, this harm may not directly impact their interests. This could explain why current strategies relying on tagging, tracking, and alerting stakeholders have failed to achieve the desired outcomes. These findings underscore the importance of focusing on the long-term impacts of retracted papers, rather than limiting attention to their short-term effects.

\begin{figure} 
	\centering
	\includegraphics[width=0.9\textwidth]{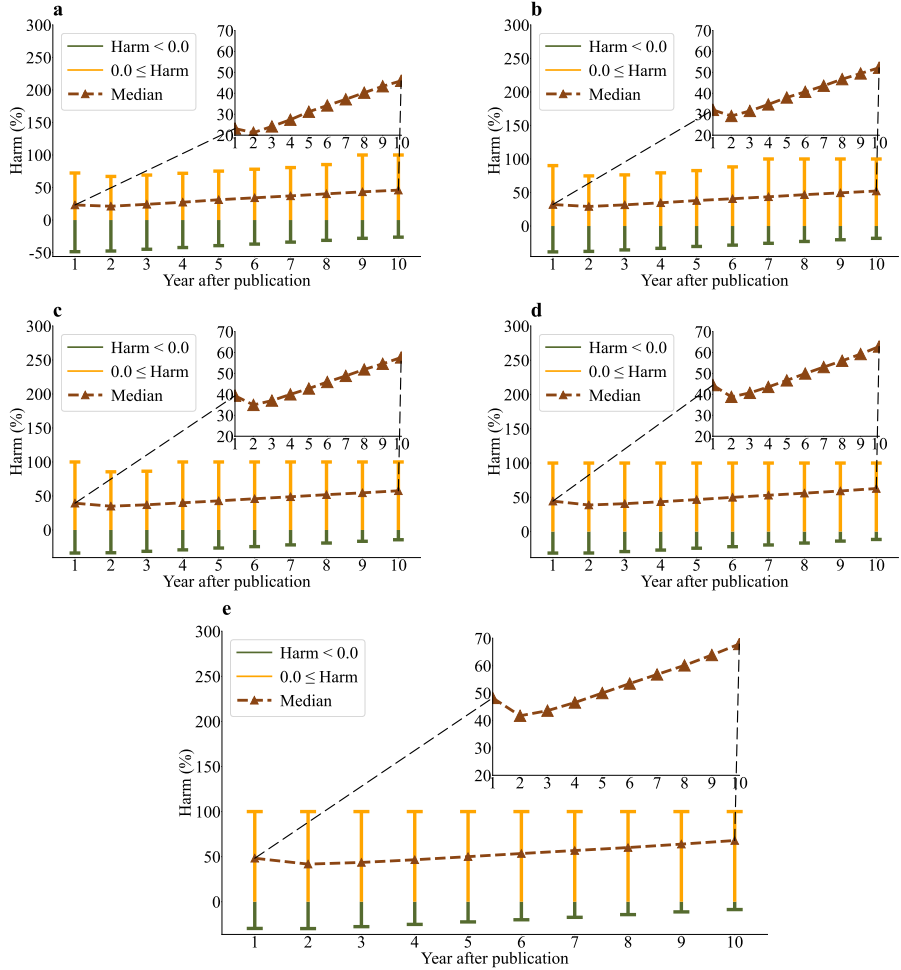} 

	\caption{\textbf{The harm of indirectly citing retracted papers.} \textbf{(a)} The median harm experienced by papers with an indirect citation distance of 2 varies over time after publication. \textbf{(b)} The median harm experienced by papers with an indirect citation distance of 3 varies over time after publication. \textbf{(c)} The median harm experienced by papers with an indirect citation distance of 4 varies over time after publication. \textbf{(d)} The median harm experienced by papers with an indirect citation distance of 5 varies over time after publication. \textbf{(e)} The median harm experienced by papers with an indirect citation distance of 6 varies over time after publication.}
	\label{fig:fig3_final}
\end{figure}

\subsection{Characteristics of the harm propagation arising from retracted papers}

To explore how the harm from retracted papers propagates, we drew upon the “six degrees of separation" theory to study articles that were connected to retracted papers via indirect citations within six steps~\cite{PhysRevX.13.021032}. As shown in Figure~\ref{fig:fig3_final}\textbf{(a)}, while the pattern of harm experienced by indirectly citing articles was similar to that of directly citing articles, the harm to indirectly citing articles was more pronounced. Specifically, within the first four years after publication, indirectly citing articles experienced a peak median harm (interquartile range) of 27.38\% (-42.30\% to 71.87\%) in the fourth year and a minimum median harm (interquartile range) of 21.22\% (-47.63\% to 67.02\%) in the second year. Compared to direct citations, indirect citations tended to reduce the visibility of retracted papers but were associated with more severe harm. As illustrated in Figures~\ref{fig:fig3_final}\textbf{(b)-(e)}, the harm to the paper indirectly citing retracted papers became increasingly significant as the distance of indirect citations grew. For articles connected via six degrees of separation, within the first four years, the median harm (interquartile range) peaked at 46.55\% (-25.15\% to 100\%) in the fourth year, while its minimum of 41.72\% (-29.80\% to 100\%) was observed in the second year. These findings suggest that as the citation distance increases, the link to the original retracted paper becomes harder to identify, yet the associated harm escalates. Detailed statistics on the harm experienced by indirectly citing articles are provided in Extended Data Table~\ref{tab:tableS2}.

Notably, with greater citation distance, the proportion of duplicate articles in the citation chain also increased significantly. To more accurately assess the impact of indirect citations, we analyzed the harm to indirectly citing articles after removing duplicates. The results revealed a pattern similar to that observed before deduplication, but the harm to deduplicated articles was even greater. Detailed data for deduplicated analyses are available in Extended Data Figure~\ref{fig:fig_s1_nd_final} and Table~\ref{tab:tableS3}.

Our findings highlight the significant and growing impact of retracted papers on indirectly citing articles, particularly as the citation network extends. These results underscore the importance of addressing the propagation of retraction-related harm, even in cases where the connection to retracted papers becomes less visible.

\begin{figure} 
	\centering
	\includegraphics[width=0.92\textwidth]{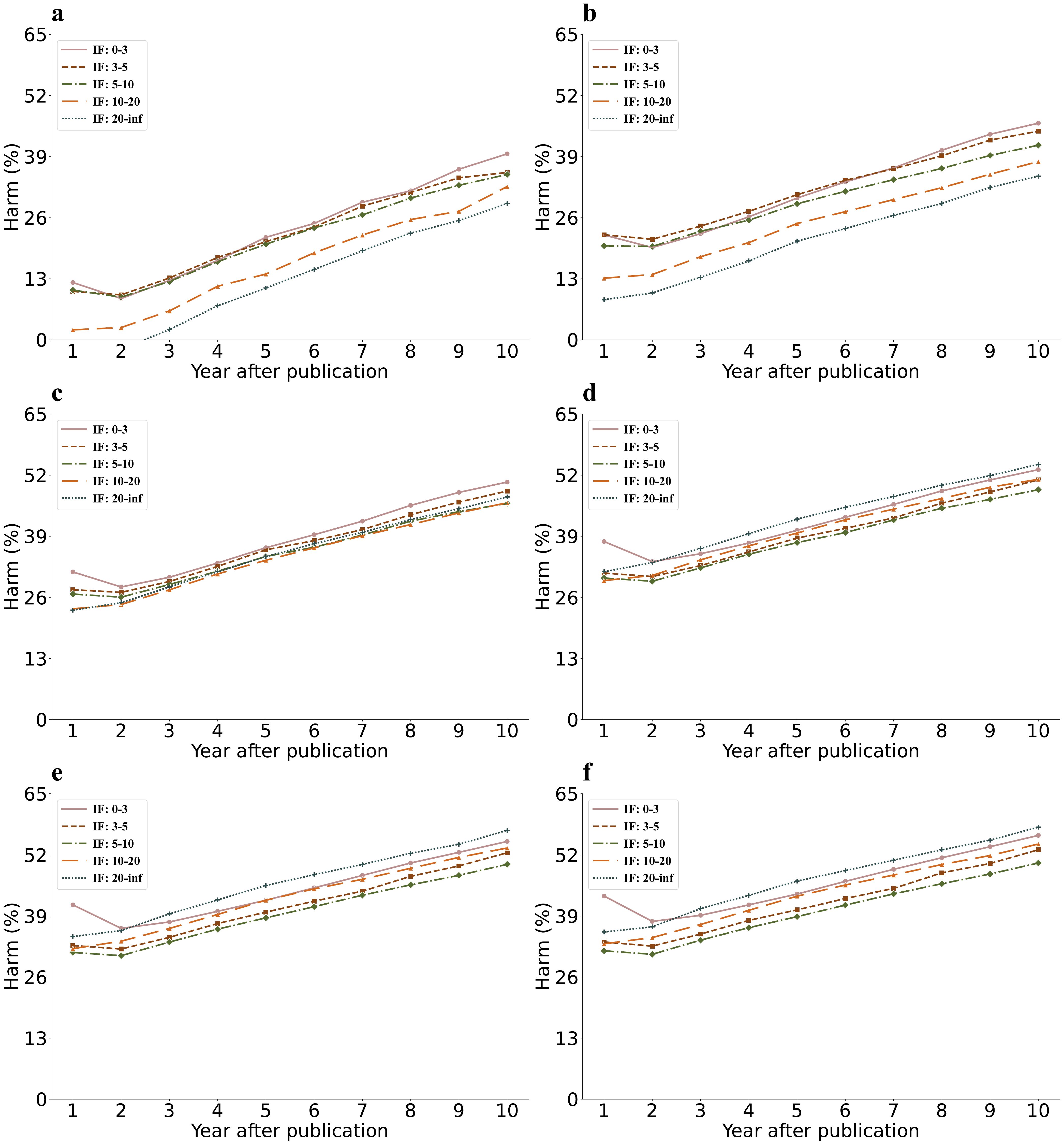} 

	\caption{\textbf{The harm experienced by papers published in journals with different impact factors (IF).} \textbf{(a)} The median harm experienced by papers directly citing retracted papers. \textbf{(b)} The median harm experienced by papers with an indirect citation distance of 2 varies over time after publication. \textbf{(c)} The median harm experienced by papers with an indirect citation distance of 3 varies over time after publication. \textbf{(d)} The median harm experienced by papers with an indirect citation distance of 4 varies over time after publication. \textbf{(e)} The median harm experienced by papers with an indirect citation distance of 5 varies over time after publication. \textbf{(f)} The median harm experienced by papers with an indirect citation distance of 6 varies over time after publication.}
	\label{fig:fig4_final}
\end{figure}

\subsection{Impact of journal prestige on the harm caused by retracted papers}
The journal impact factor (IF) is widely used to quantify a journal's influence. Although it does not fully capture the nuances of a journal’s impact, it remains the most widely accepted metric in academic publishing~\cite{wang2013quantifying}. 
To investigate how journal impact factors influence the harm caused by retracted papers, we utilized the SciSciNet dataset (99,845,112 papers) to match all articles with their respective journals and analyzed papers published across journals with varying IFs. As shown in Figure~\ref{fig:fig4_final}\textbf{(a)}, articles published in journals with an IF below 10 experienced significantly greater harm compared to those in journals with an IF above 10. Remarkably, for articles published in journals with an IF exceeding 20, the median harm during the first few years after publication was minimal and could even be negative.


A closer look at Figures~\ref{fig:fig4_final}\textbf{(b)-(f)} reveals that, while the overall harm patterns are consistent across all IF categories, notable differences appear when examining citation distances of two steps or fewer. Articles published in journals with an IF above 10 suffered significantly less harm than those published in journals with an IF below 10. In contrast, for citation distances greater than two, the impact factor of the publishing journal did not exhibit a statistically significant effect on the harm experienced by citing articles. Detailed data is available in Extended Data Table~\ref{tab:tableS4}.

These results suggest that higher-impact journals may provide a protective effect against retraction-related harm for directly citing articles and those within close indirect citation networks. However, this protective influence diminishes as the citation distance increases. 


\begin{figure} 
	\centering
	\includegraphics[width=0.85\textwidth]{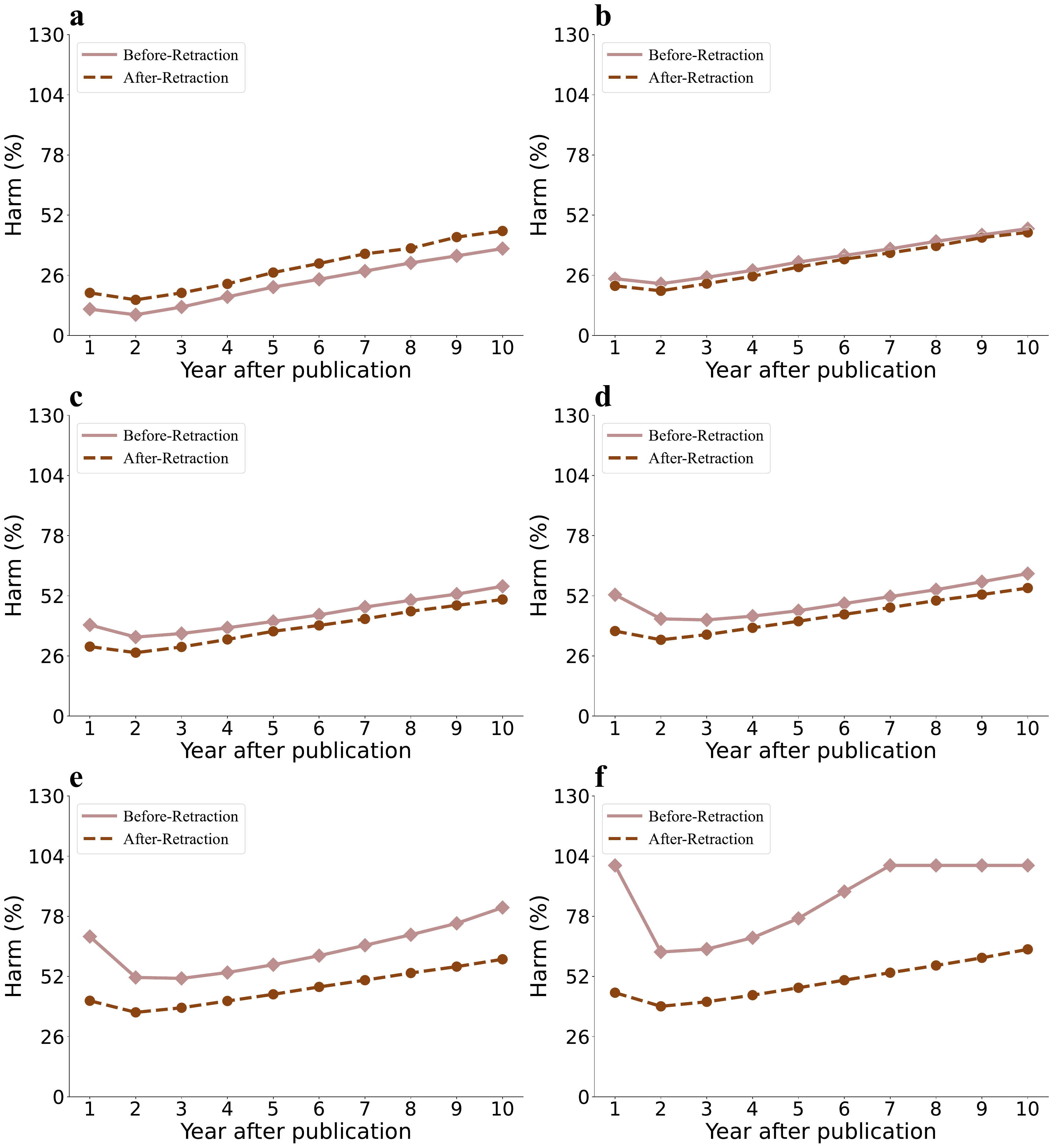} 

	\caption{\textbf{The harm experienced by citing retracted papers before and after retraction.} \textbf{(a)} The median harm experienced by papers directly citing retracted papers before and after retraction. \textbf{(b)} The median harm experienced by papers with an indirect citation distance of 2 before and after retraction varies over time after their publication. \textbf{(c)} The median harm experienced by papers with an indirect citation distance of 3 before and after retraction varies over time after their publication. \textbf{(d)} The median harm experienced by papers with an indirect citation distance of 4 before and after retraction varies over time after their publication. \textbf{(e)} The median harm experienced by papers with an indirect citation distance of 5 before and after retraction varies over time after their publication. \textbf{(f)} The median harm experienced by papers with an indirect citation distance of 6 before and after retraction varies over time after their publication.}
	\label{fig:fig5_final}
\end{figure}

\subsection{Characteristics of harm caused by retracted papers before and after retraction}
Retraction has become an important measure to mitigate the negative impacts of problematic papers and is considered to play a critical remedial and deterrent role in curbing the publication and dissemination of such papers. To investigate the effect of retraction on the spread of academic harm, we analyzed citation patterns before and after retraction. As shown in Figure~\ref{fig:fig5_final}\textbf{(a)}, the harm experienced by papers directly citing retracted articles increased by citations made after the retraction. Prior to retraction, papers citing the retracted papers exhibited a median maximum harm (interquartile range) of 16.67\% (-63.94\% to 65.39\%) within the first four years after publication. After Retraction, the median maximum harm (interquartile range) increased to 22.28\% (-59.67\% to 72.98\%).

In contrast, as illustrated in Figrue~\ref{fig:fig5_final}\textbf{(b)-(d)}, the harm to papers with indirect citation distances of 2–4 was somewhat mitigated by retraction, though the effect was not pronounced. Furthermore, as depicted in Figure~\ref{fig:fig5_final}\textbf{(e)-(f)}, retraction had a significant effect in reducing the harm to papers with indirect citation distances of 5 and 6. However, papers citing retracted articles after their retraction continued to experience considerable harm. Detailed data is available in Extended Data Table~\ref{tab:tableS5}.

These findings suggest that retraction, by potentially amplifying the early harm of retracted papers, may help draw the attention of stakeholders, thereby facilitating intervention. Additionally, retraction has a positive effect on indirectly citing papers, with this positive impact increasing as the citation distance grows. However, it is important to note that even with these positive effects, the harm experienced by these papers remains significant.

\section{Discussion}\label{sec3}

In this study, we propose an innovative measurement framework that quantifies the impact of retracted papers on the scientific community for the first time. Through a systematic analysis of the harms caused by retracted papers, we identify an “attention escape" mechanism that enables retracted papers to persistently inflict significant negative effects on the scientific community. Specifically, our findings reveal the following characteristics of the harm caused by retracted papers: short-term effects are insignificant, but long-term effects are substantial; the impact is smaller on directly cited papers but significant on indirectly cited ones; the harm is less pronounced in high-impact journals but more significant in lower-impact journals. Our findings provide new insights into the development of more effective strategies for addressing the issues posed by retracted papers and lay the groundwork for advancing research in this field.

This study provides the first quantitative analysis of the harm caused by retracted papers and the mechanisms underlying their dissemination, uncovering specific patterns and potential drivers of their spread. While, as expected, retraction does impact the harm caused by these papers, the effects are limited. Relying solely on punitive measures, such as retraction or marking retracted citations, to curb their dissemination remains challenging. Our identification of the “attention escape" mechanism in the harm caused by retracted papers prompts a reevaluation of strategies, suggesting a shift from penalizing stakeholders to encouraging proactive accountability through highlighting reputational and institutional risks. Our findings will drive the field to shift from merely issuing warnings to stakeholders toward awakening their intrinsic sense of responsibility. This transformation will pave the way for more comprehensive and effective approaches, strengthening research integrity and fostering a greater sense of accountability across the scientific community.

Despite the limited success or uncertain impact of many current efforts to mitigate the harm caused by retracted papers and prevent their dissemination, our findings offer a promising path for improving these strategies. Existing approaches primarily focus on tagging and tracking papers that directly cite retracted work. However, our research highlights a critical disparity: while the harm to directly citing papers is relatively limited, the impact on indirectly citing papers is significantly greater. This distinction underscores the urgent need to broaden the scope of current tools and strategies to include the tracking of indirect citations, as these papers are more likely to experience substantial harm caused by retracted works. By emphasizing the risks associated with indirect citations, targeted strategies and tools can effectively raise stakeholder awareness and foster more proactive engagement. Such measures not only address the secondary dissemination of harm but also provide stakeholders with meaningful opportunities to take corrective actions. This perspective bridges the gap in current research, not only supporting the improvement of governance over the harm caused by retracted papers but also paving new pathways for advancing research in this critical area.

Our findings indicate that the harm caused by retracted papers and their dissemination is closely linked to entrenched “norms" in academia. Specifically, the excessive emphasis on recent research while neglecting long-term contributions, combined with the reliance on journal impact factors as primary evaluation metrics, prevents stakeholders from effectively addressing the harm caused by retracted papers. In the absence of punitive measures, addressing these enduring “norms" within the academic system becomes critically important. Our study connects the harm caused by problematic papers to systemic challenges within the academic recognition framework, offering an additional perspective and direction for addressing this issue.

Our study has certain limitations. The forms of harm are not limited to paper citations; they may also include factors such as the likelihood of publication, funding acquisition, and reputation. However, quantifying aspects such as reputation and funding is challenging, and data availability is limited. Meanwhile, the publication and citation of papers are widely recognized as the “currency" in the academic community and are closely linked to other factors like reputation and funding acquisition~\cite{cabanac2024chain}. Therefore, analyzing the citation patterns of papers can largely reflect the harm caused by retracted papers. Additionally, due to the lack of data on the paper submission process, we were unable to directly connect publication with the harm caused by retracted papers or explore the potential harm associated with the publication process. However, citation count remains crucial in academic evaluation, effectively revealing both the harm caused by retracted papers and the propagation of that harm. Our findings have broad policy implications, offering valuable insights for policy development. Citation-based evaluation mechanisms are currently widespread across various stakeholders (e.g., institutions, authors, journals), influencing reward programs, research funding allocation, award decisions, and even salary and bonus structures, as well as journal rankings~\cite{fuyuno2006cash}.

Our study provides the first direct quantitative evidence of the destructive mechanisms of retracted papers, enhancing the scientific community's understanding of their harmful impacts. It highlights significant research gaps in addressing the detrimental effects of retracted papers. Looking forward, we aim to leverage our findings to develop effective tools and pursue further in-depth research. Specifically, we plan to create a system capable of real-time detection and tracking of retracted papers while quantitatively analyzing the potential harm to both directly and indirectly citing papers. This system will offer valuable insights for both published works and manuscripts in preparation, aiming to contain the spread of harmful impacts and prevent the further dissemination of flawed research. Additionally, we will explore the feasibility of various identification mechanisms to redirect stakeholders' attention toward mitigating the damage caused by problematic papers.

\section{Methods}




\subsection{Data sources and preparation}

\textbf{Semantic Scholar Database.}
The Semantic Scholar Database (\textit{SSD}) \cite{kinney2023semantic} covers a wide range of publication records, authors, institutions, entities and various interactions between them (e.g., authorships, citations). 
Here we use the edition released on January 4, 2024 by \textit{SSD}, in total covering 215,943,888 publication records and 2,296,048,817 citation records.
The database contains a variety of paper metadata, such as the title, DOI, authors, journal, field, publication date, citation count, etc. 
Prior to analysis, the \textit{SSD} underwent a series of preprocessing steps to ensure data quality and suitability for the intended research. 
Initially, publications lacking information regarding their publication year were removed, as this information is crucial for determining the starting point of citation analysis. Subsequently, publications published before 1900 were excluded to focus on more contemporary scholarly work. Furthermore, deduplication of publications based on DOI was performed. For publications with identical DOIs, those with a higher citation count were prioritized; in cases of equal citation counts, the publication with a greater number of references was retained, resulting in 210,309,367 publications. 

To ensure a minimum citation window of ten years for analysis, the dataset was further filtered to include only publications with a publication year of 2013 or earlier, resulting in 133,117,934 publications. Finally, to facilitate comparisons with publications in the same journal and field of study, records lacking either journal information or field of study information were removed, yielding a final dataset of 29,752,635 publications and 360,775,187 citation records.

\vspace{0.15in}

\noindent\textbf{Retraction Watch Database.} 
We obtained the set of retracted articles from Retraction Watch \cite{RetractionWatch}. This database encompasses 49,721 papers retracted as of 2024, with each record containing metadata including the title, journal, publication date, and retraction date. 
The retraction database underwent initial preprocessing steps to ensure data integrity. Initially, records with a null value for DOI were removed, as this field is essential for identifying the retracted publication. Subsequently, duplicate records identified by identical DOI values were eliminated, and the record with the latest retraction publication date was retained. Finally, the processed retraction dataset was matched with the \textit{SSD} via the DOI to link retraction information with publication metadata. This preprocessing resulted in a final dataset of 39,449 retracted publications.

\vspace{0.15in}

\noindent\textbf{SciSciNet Database.} 
The SciSciNet dataset \cite{lin2023sciscinet}, an extension of the Microsoft Academic Graph (MAG) offering comprehensive author, affiliations, journal metadata and linkage information among these, contained 136,726,948 publications.
To facilitate effective matching with the \textit{SSD}, this dataset underwent preprocessing. This involved the removal of publication records lacking DOI and the subsequent deduplication of entries based on DOI. The deduplication strategy prioritized retaining records with higher citation counts and, secondarily, a greater number of references. Following these steps, the dataset was reduced to 99,845,112 publications. This database serves as a valuable resource for augmenting the information contained within the \textit{SSD}.

\subsection{Quantification of harm}

To quantify the potential harm associated with retracted papers, we focus on the papers that directly or indirectly cite these retractions (hereafter referred to as papers '$C$'). We define this harm as manifesting in the citation patterns of these papers $C$ . Specifically, we operationalize harm as the ratio of citations received by a paper $C$ compared to the average citations received by a set of comparable papers (hereafter referred to as papers '$D$'). These papers $D$ are defined as those published in the same venue, around the same time, and within the same field as the paper $C$.

\vspace{0.15in}

\noindent\textbf{Citation Distance.}
We define different levels of citation relationships. Let $C_1$ represent papers that directly cite retracted papers, $C_2$ represent papers that cite $C_1$ (thus indirectly citing retracted papers), and so on, forming a citation chain $C_1, C_2, ..., C_n$ where $n$ represents the citation depth level. 

\vspace{0.15in}

\noindent\textbf{Harm Measurement.} 
For each citing $pr_c$ belonging to the set of citing papers $C$, we compute a harm based on its citation patterns relative to a comparator group $D$. Given a citing $pr_c$ published in venue $v_c$ in year $y_c$, and associated with a set of fields of study $F_c = \{f_{c1}, f_{c2}, ..., f_{cm}\}$, the comparator set $pr_d$ (denoted as $D(c)$) is constructed by selecting papers that satisfy the following criteria:
\begin{itemize}
    \item \textbf{Same Venue:} $\text{venue}(pr_d) = v_c$
    \item \textbf{Temporally Proximate Publication Year:} $\text{year}(pr_d) \in \{y_c - 1, y_c, y_c + 1\}$
    \item \textbf{Matching Field of Study:} $\text{field}(pr_d) \cap F_c \neq \emptyset$
\end{itemize}

Formally, the comparator set \(D(c)\) is defined as:
\begin{equation}
    D(c) = \{ pr_d \mid \text{venue}(pr_d) = v_c \land \text{year}(pr_d) \in \{y_c - 1, y_c, y_c + 1\} \land \text{field}(pr_d) \cap F_c \neq \emptyset \}
\end{equation}
where venue, year, and field information are obtained from the \textit{SSD}.

Next, we calculate aggregate citation metrics for the comparator set $D(c)$:
\begin{align}
    prs_d\_ct &= \sum_{pr_d \in D(c)} \text{citation\_count}(pr_d) \\
    prs_d\_cy[k] &= \sum_{pr_d \in D(c)} \text{citations\_year\_k}(pr_d), \quad k \in \{1, 2, ..., 10\}
\end{align}
where $prs_d\_ct$ represents the total citations for the comparator set $D(c)$, $\text{citation\_count}(pr_d)$ is the total citation count for each paper, $prs\_d\_cy[k]$ represents the yearly citations for the comparator set $D(c)$ in the $k$-th year after publication, and $\text{citations\_year\_k}(pr_d)$ is the number of citations received by a paper in the $k$-th year after its publication. Both $\text{citation\_count}(pr_d)$ and $\text{citations\_year\_k}(pr_d)$ are derived from the \textit{SSD}. The total citation count for each paper is obtained from the \textit{SSD} and the citations for each year after publication are derived from the citation records within the \textit{SSD}.


Let $n(d)$ denote the number of papers in the comparator set $D(c)$. The harm for \(pr_c\) is then computed by comparing its citation metrics to the average citation metrics of its comparator set \(D(c)\):
\begin{align}
    harm\_ct(pr_c) &= 1 - \frac{\text{citation\_count}(pr_c)}{\frac{prs_d\_ct}{n(d)}}, \quad \text{if } n(d) > 0 \text{ and } prs_d\_ct > 0 \\
    harm\_cy[k](pr_c) &= 1 - \frac{\text{citations\_year\_k}(pr_c)}{\frac{pr_d\_cy[k]}{n(d)}}, \quad \text{if } n(d) > 0 \text{ and } prs_d\_cy[k] > 0
\end{align}
where $harm\_ct(pr_c)$ represents the harm for total citations for paper $c$, $harm\_cy[k](pr_c)$ represents the harm for yearly citations for paper $c$ in year $k$.

Finally, for each citing paper \(pr_c\), we construct a harm vector \(h(pr_c)\):
\begin{align}
h(pr_c) &= [h_0, h_1, h_2, ..., h_{10}]
\end{align}
where,
\begin{align}
    h_0 &= harm\_ct(pr_c) \nonumber \\
    h_k &= harm\_cy[k](pr_c), \quad k \in \{1, 2, ..., 10\}
\end{align}

This $h(pr_c)$ represents the computed harm for \(pr_c\) based on its total and yearly citation patterns relative to its comparator group.

\subsection{Statistical analysis}

Following the computation of the harm for each paper $C$, as defined in the previous section, we conduct a statistical analysis to investigate the trends and patterns of this metric across different research domains and in aggregate. This analysis is divided into three main parts.


\vspace{0.15in}

\noindent\textbf{Field-Based Analysis of Harm.} 
We analyze the overall and field-specific trends of the harm over a 10-year window following each paper's publication. For both the entire corpus and individual fields, we compute the quartile statistics (Q1, Q2, Q3) of the harm distribution:
\begin{align}
    H_{all} &= \{Q1, Q2, Q3\} \text{ of } \{h(p) \mid p \in P\} \\
    H_f &= \{Q1, Q2, Q3\} \text{ of } \{h(p) \mid p \in P_f\}, \quad \forall f \in F
\end{align}
where $P$ represents all papers in the corpus, $P_f$ denotes papers in field $f$, $F$ is the set of all fields, and $h(p)$ is the harm of paper $p$. Note that papers with multiple field affiliations contribute to multiple $P_f$ sets.

\vspace{0.15in}

\noindent\textbf{Impact Factor-Based Analysis of Harm.}
We further analyze the harm distribution across different journal impact factor (IF) ranges. Papers are categorized into five groups based on their venue's impact factor: $[0,3), [3,5), [5,10), [10,20)$, and $[20,\infty)$. Similar to the field-based analysis, we compute quartile statistics for each IF group:
\begin{align}
    IF_{ranges} &= \{[0,3), [3,5), [5,10), [10,20), [20,\infty)\} \\
    H_{if} &= \{Q1, Q2, Q3\} \text{ of } \{h(p) \mid p \in P_{if}\}, \quad \forall if \in IF_{ranges}
\end{align}
where $P_{if}$ denotes the papers whose publishing venues have impact factors in range $if$.

\vspace{0.15in}

\noindent\textbf{Harm Analysis of Papers Citing Retracted Papers Before and After Retraction.}
We categorize papers based on the date they cited (directly or indirectly) retracted papers relative to the retraction date. Let $r$ be a retracted paper and $c$ be a paper that cites $r$, either directly or indirectly through citation chains. We divide papers into two groups according to whether their citations occurred before or after the retraction of $r$. Similar to the previous analyses, we compute quartile statistics for two groups of papers. For papers citing multiple retracted papers, we use the earliest retraction date as the reference point.
\begin{align}
    H_{pre} &= \{Q1, Q2, Q3\} \text{ of } \{h(c) \mid \text{cite\_time}(c, r) < \min_{r \in R_c}\{\text{retraction\_time}(r)\}\} \\
    H_{post} &= \{Q1, Q2, Q3\} \text{ of } \{h(c) \mid \text{cite\_time}(c, r) \geq \min_{r \in R_c}\{\text{retraction\_time}(r)\}\}
\end{align}
where $\text{cite\_time}(c, r)$ represents the date paper $c$ cited (directly or indirectly) the retracted paper $r$, $\text{retraction\_time}(r)$ is the retraction date of $r$, and $R_c$ is the set of all retracted papers directly or indirectly cited by paper $c$.

\section*{Declarations}

\noindent\textbf{Funding.}
We acknowledge supports from the Innovation Funding of ICT, CAS under Grant No. E461070.

\vspace{0.15in}

\noindent\textbf{Competing interests.}
The authors declare no conflicts of interest and no competing interests. 

\vspace{0.15in}

\noindent\textbf{Data availability.}
All data involved in this study are publicly available via GitHub at \url{https://github.com/Garfyyy/quantifying-retraction-harm}.

\vspace{0.15in}

\noindent\textbf{Code availability.}
The code to reproduce the main results in the figures from the aggregated data is publicly available on GitHub via \url{https://github.com/Garfyyy/quantifying-retraction-harm}.

\vspace{0.15in}

\noindent\textbf{Author contributions.}
Y.H. conceptualized this study, formulated the designs, conceived the experiments, and wrote the manuscript. J.Z. implemented the algorithms, collected and analyzed the data. D.C., Z.Y., B.X., Q.L, W.J., L.Ma., and S.T. analyzed the data. T.H., Z.Z., W.G. and J.Z. conceptualized this study, directed the project, and revised the manuscript. All authors have read and approved the final manuscript.

\bibliography{sn-article}

\makeatletter
\renewcommand{\fnum@figure}{\textbf{Extended Data Fig. \thefigure}}
\renewcommand{\fnum@table}{\textbf{Extended Data Table \thetable}}
\makeatother

\setcounter{figure}{0} 

\begin{figure}[ht] 
	\centering
	\includegraphics[width=0.9\textwidth]{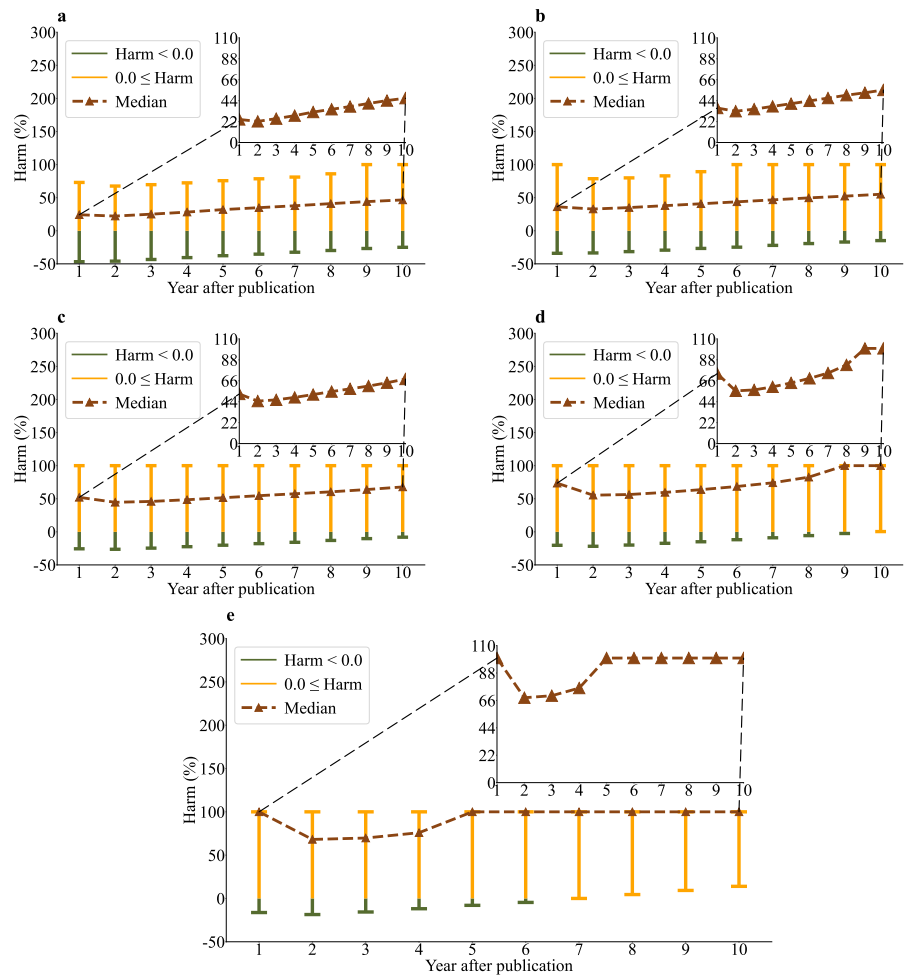} 
	\caption{\textbf{The harm of citing retracted papers, with no duplicate statistics across different citation distances.} \textbf{(a)} The median harm experienced by papers with an indirect citation distance of 2 varies over time after publication. \textbf{(b)} The median harm experienced by papers with an indirect citation distance of 3 varies over time after publication. \textbf{(c)} The median harm experienced by papers with an indirect citation distance of 4 varies over time after publication. \textbf{(d)} The median harm experienced by papers with an indirect citation distance of 5 varies over time after publication. \textbf{(e)} The median harm experienced by papers with an indirect citation distance of 6 varies over time after publication.}
	\label{fig:fig_s1_nd_final} 
\end{figure}

\begin{sidewaystable}[]
\centering
\caption{The median harm experienced by papers directly citing retracted articles across different fields varies over time after their publication (Part 1).}
\label{tab:tableS1}
\begin{tabularx}{0.95\columnwidth}{@{\extracolsep\fill}
w{l}{1.5cm}
w{c}{0.4cm}
ccccc
}
\toprule
Field &
  Nums &
  Y1 Harm (IQR) &
  Y2 Harm (IQR) &
  Y3 Harm (IQR) &
  Y4 Harm (IQR) &
  Y5 Harm (IQR) \\ 
  \midrule
All &
  119644 &
  \begin{tabular}[c]{@{}c@{}}13.08\\ (-69.11$\sim$67.57)\end{tabular} &
  \begin{tabular}[c]{@{}c@{}}10.68\\ (-68.71$\sim$62.46)\end{tabular} &
  \begin{tabular}[c]{@{}c@{}}13.99\\ (-65.50$\sim$64.97)\end{tabular} &
  \begin{tabular}[c]{@{}c@{}}18.67\\ (-62.92$\sim$68.09)\end{tabular} &
  \begin{tabular}[c]{@{}c@{}}22.84\\ (-57.55$\sim$71.98)\end{tabular} \\
Agricultural &
  1439 &
  \begin{tabular}[c]{@{}c@{}}16.38\\ (-55.31$\sim$74.47)\end{tabular} &
  \begin{tabular}[c]{@{}c@{}}8.27\\ (-62.50$\sim$64.40)\end{tabular} &
  \begin{tabular}[c]{@{}c@{}}12.26\\ (-64.22$\sim$62.94)\end{tabular} &
  \begin{tabular}[c]{@{}c@{}}11.60\\ (-63.43$\sim$64.10)\end{tabular} &
  \begin{tabular}[c]{@{}c@{}}18.18\\ (-62.82$\sim$68.42)\end{tabular} \\
Art &
  31 &
  \begin{tabular}[c]{@{}c@{}}41.18\\ (-17.39$\sim$100.00)\end{tabular} &
  \begin{tabular}[c]{@{}c@{}}45.45\\ (-88.14$\sim$100.00)\end{tabular} &
  \begin{tabular}[c]{@{}c@{}}33.33\\ (-91.15$\sim$100.00)\end{tabular} &
  \begin{tabular}[c]{@{}c@{}}0.00\\ (-87.43$\sim$100.00)\end{tabular} &
  \begin{tabular}[c]{@{}c@{}}10.24\\ (-150.00$\sim$85.00)\end{tabular} \\
Biology &
  57649 &
  \begin{tabular}[c]{@{}c@{}}9.41\\ (-67.83$\sim$59.57)\end{tabular} &
  \begin{tabular}[c]{@{}c@{}}7.97\\ (-67.21$\sim$56.02)\end{tabular} &
  \begin{tabular}[c]{@{}c@{}}11.49\\ (-63.32$\sim$59.61)\end{tabular} &
  \begin{tabular}[c]{@{}c@{}}16.21\\ (-60.53$\sim$62.67)\end{tabular} &
  \begin{tabular}[c]{@{}c@{}}20.73\\ (-54.93$\sim$66.38)\end{tabular} \\
\bottomrule
\end{tabularx}%

\vspace{0.3cm}

\centering
\begin{tabularx}{0.95\columnwidth}{@{\extracolsep\fill}
w{l}{1.5cm}
w{c}{0.4cm}
ccccc
}
\toprule
 &
   &
  Y6 Harm (IQR) &
  Y7 Harm (IQR) &
  Y8 Harm (IQR) &
  Y9 Harm (IQR) &
  Y10 Harm (IQR) \\
  \midrule
 &
   &
  \begin{tabular}[c]{@{}c@{}}26.49\\ (-55.13$\sim$75.61)\end{tabular} &
  \begin{tabular}[c]{@{}c@{}}30.43\\ (-51.75$\sim$78.46)\end{tabular} &
  \begin{tabular}[c]{@{}c@{}}33.44\\ (-48.23$\sim$81.69)\end{tabular} &
  \begin{tabular}[c]{@{}c@{}}37.21\\ (-44.10$\sim$89.09)\end{tabular} &
  \begin{tabular}[c]{@{}c@{}}40.08\\ (-41.39$\sim$100.00)\end{tabular} \\
 &
   &
  \begin{tabular}[c]{@{}c@{}}20.59\\ (-60.17$\sim$72.26)\end{tabular} &
  \begin{tabular}[c]{@{}c@{}}26.09\\ (-48.48$\sim$75.91)\end{tabular} &
  \begin{tabular}[c]{@{}c@{}}27.44\\ (-53.12$\sim$77.33)\end{tabular} &
  \begin{tabular}[c]{@{}c@{}}31.66\\ (-43.10$\sim$80.24)\end{tabular} &
  \begin{tabular}[c]{@{}c@{}}32.33\\ (-43.69$\sim$86.65)\end{tabular} \\
 &
   &
  \begin{tabular}[c]{@{}c@{}}14.29\\ (-72.04$\sim$100.00)\end{tabular} &
  \begin{tabular}[c]{@{}c@{}}46.19\\ (-94.59$\sim$100.00)\end{tabular} &
  \begin{tabular}[c]{@{}c@{}}40.59\\ (-200.00$\sim$100.00)\end{tabular} &
  \begin{tabular}[c]{@{}c@{}}56.64\\ (-96.63$\sim$100.00)\end{tabular} &
  \begin{tabular}[c]{@{}c@{}}100.00\\ (-18.45$\sim$100.00)\end{tabular} \\
 &
   &
  \begin{tabular}[c]{@{}c@{}}23.78\\ (-52.73$\sim$69.45)\end{tabular} &
  \begin{tabular}[c]{@{}c@{}}27.79\\ (-50.51$\sim$72.48)\end{tabular} &
  \begin{tabular}[c]{@{}c@{}}30.97\\ (-46.59$\sim$75.12)\end{tabular} &
  \begin{tabular}[c]{@{}c@{}}33.71\\ (-42.90$\sim$77.07)\end{tabular} &
  \begin{tabular}[c]{@{}c@{}}36.38\\ (-40.62$\sim$80.80)\end{tabular} \\
\bottomrule
\end{tabularx}
\end{sidewaystable}

\addtocounter{table}{-1}
\begin{sidewaystable}[]
\centering
\caption{The median harm experienced by papers directly citing retracted articles across different fields varies over time after their publication (Part 2).}
\label{tab:tableS1c}
\begin{tabularx}{0.95\columnwidth}{@{\extracolsep\fill}
w{l}{1.5cm}
w{c}{0.4cm}
ccccc
}
\toprule
Field &
  Nums &
  Y1 Harm (IQR) &
  Y2 Harm (IQR) &
  Y3 Harm (IQR) &
  Y4 Harm (IQR) &
  Y5 Harm (IQR) \\ 
  \midrule
Geology &
  86 &
  \begin{tabular}[c]{@{}c@{}}37.72\\ (-18.75$\sim$78.54)\end{tabular} &
  \begin{tabular}[c]{@{}c@{}}34.94\\ (-44.04$\sim$81.95)\end{tabular} &
  \begin{tabular}[c]{@{}c@{}}38.12\\ (-36.41$\sim$79.59)\end{tabular} &
  \begin{tabular}[c]{@{}c@{}}27.49\\ (-19.33$\sim$88.46)\end{tabular} &
  \begin{tabular}[c]{@{}c@{}}35.94\\ (-35.81$\sim$87.92)\end{tabular} \\
History &
  96 &
  \begin{tabular}[c]{@{}c@{}}44.25\\ (-50.00$\sim$100.00)\end{tabular} &
  \begin{tabular}[c]{@{}c@{}}47.06\\ (-58.89$\sim$100.00)\end{tabular} &
  \begin{tabular}[c]{@{}c@{}}43.42\\ (-62.50$\sim$100.00)\end{tabular} &
  \begin{tabular}[c]{@{}c@{}}55.77\\ (-59.34$\sim$100.00)\end{tabular} &
  \begin{tabular}[c]{@{}c@{}}59.62\\ (-19.51$\sim$100.00)\end{tabular} \\
Law &
  127 &
  \begin{tabular}[c]{@{}c@{}}80.01\\ (0.00$\sim$100.00)\end{tabular} &
  \begin{tabular}[c]{@{}c@{}}72.36\\ (9.68$\sim$100.00)\end{tabular} &
  \begin{tabular}[c]{@{}c@{}}71.32\\ (-0.41$\sim$100.00)\end{tabular} &
  \begin{tabular}[c]{@{}c@{}}76.13\\ (7.85$\sim$100.00)\end{tabular} &
  \begin{tabular}[c]{@{}c@{}}89.42\\ (18.59$\sim$100.00)\end{tabular} \\
Linguistics &
  42 &
  \begin{tabular}[c]{@{}c@{}}25.00\\ (-45.45$\sim$100.00)\end{tabular} &
  \begin{tabular}[c]{@{}c@{}}6.68\\ (-67.46$\sim$71.76)\end{tabular} &
  \begin{tabular}[c]{@{}c@{}}28.39\\ (-60.00$\sim$100.00)\end{tabular} &
  \begin{tabular}[c]{@{}c@{}}35.31\\ (-41.43$\sim$100.00)\end{tabular} &
  \begin{tabular}[c]{@{}c@{}}50.21\\ (-33.60$\sim$100.00)\end{tabular} \\
\bottomrule
\end{tabularx}

\vspace{0.3cm}

\centering
\begin{tabularx}{0.95\columnwidth}{@{\extracolsep\fill}
w{l}{1.5cm}
w{c}{0.4cm}
ccccc
}
\toprule
 &
   &
  Y6 Harm (IQR) &
  Y7 Harm (IQR) &
  Y8 Harm (IQR) &
  Y9 Harm (IQR) &
  Y10 Harm (IQR) \\ 
  \midrule
 &
   &
  \begin{tabular}[c]{@{}c@{}}36.62\\ (-25.61$\sim$79.48)\end{tabular} &
  \begin{tabular}[c]{@{}c@{}}36.74\\ (-40.79$\sim$80.02)\end{tabular} &
  \begin{tabular}[c]{@{}c@{}}29.87\\ (-40.24$\sim$88.24)\end{tabular} &
  \begin{tabular}[c]{@{}c@{}}19.64\\ (-37.21$\sim$87.21)\end{tabular} &
  \begin{tabular}[c]{@{}c@{}}45.68\\ (-8.67$\sim$100.00)\end{tabular} \\
 &
   &
  \begin{tabular}[c]{@{}c@{}}54.40\\ (-23.27$\sim$100.00)\end{tabular} &
  \begin{tabular}[c]{@{}c@{}}65.82\\ (-39.28$\sim$100.00)\end{tabular} &
  \begin{tabular}[c]{@{}c@{}}60.09\\ (-63.32$\sim$100.00)\end{tabular} &
  \begin{tabular}[c]{@{}c@{}}79.49\\ (-28.95$\sim$100.00)\end{tabular} &
  \begin{tabular}[c]{@{}c@{}}76.90\\ (-35.44$\sim$100.00)\end{tabular} \\
 &
   &
  \begin{tabular}[c]{@{}c@{}}100.00\\ (42.21$\sim$100.00)\end{tabular} &
  \begin{tabular}[c]{@{}c@{}}100.00\\ (37.00$\sim$100.00)\end{tabular} &
  \begin{tabular}[c]{@{}c@{}}100.00\\ (30.29$\sim$100.00)\end{tabular} &
  \begin{tabular}[c]{@{}c@{}}100.00\\ (43.86$\sim$100.00)\end{tabular} &
  \begin{tabular}[c]{@{}c@{}}100.00\\ (42.14$\sim$100.00)\end{tabular} \\
 &
   &
  \begin{tabular}[c]{@{}c@{}}57.91\\ (-22.28$\sim$100.00)\end{tabular} &
  \begin{tabular}[c]{@{}c@{}}56.80\\ (-32.35$\sim$100.00)\end{tabular} &
  \begin{tabular}[c]{@{}c@{}}51.92\\ (-1.37$\sim$100.00)\end{tabular} &
  \begin{tabular}[c]{@{}c@{}}67.47\\ (-14.58$\sim$100.00)\end{tabular} &
  \begin{tabular}[c]{@{}c@{}}66.80\\ (-13.04$\sim$100.00)\end{tabular} \\
\bottomrule
\end{tabularx}
\end{sidewaystable}

\begin{sidewaystable}[]
\scriptsize
\centering
\caption{Whether top 30 universities emphasize recent research outputs in promotion decisions}
\label{tab:tableS6}
\begin{tabularx}{1.05\textheight}{@{\extracolsep\fill}
w{c}{0.4cm}
w{l}{5cm}
w{c}{1.1cm}
w{c}{1.5cm}
X
}
\toprule
RANK &
  Institution Name &
  \makecell{ISO3166\\Code} &
  \makecell{Promotion\\Information} &
  Url \\ \midrule
1 &
  Massachusetts Institute of Technology &
  US &
  yes &
  \url{https://policies.mit.edu/policies-procedures/30-faculty-appointment-promotion-and-tenure-guidelines} \\
2 &
  University of Cambridge &
  GB &
  yes &
  \url{https://www.acp.hr.admin.cam.ac.uk/application-promotion} \\
3 &
  University of Oxford &
  GB &
  yes &
  \url{https://hr.admin.ox.ac.uk/guide-to-the-appointment-associate-professors} \\
4 &
  Harvard University &
  US &
  yes &
  \url{https://academic-appointments.fas.harvard.edu/internal-promotion-tenured-professor-tenure-track-position} \\
5 &
  Stanford University &
  US &
  yes &
  \url{https://facultyhandbook.stanford.edu/index/chapter-2-appointments-reappointments-and-promotions-professoriate} \\
6 &
  Imperial College London &
  GB &
  yes &
  \url{https://www.imperial.ac.uk/human-resources/recruitment-and-promotions/promotions/academic-promotions} \\
7 &
  ETH Zurich &
  CH &
  yes &
  \url{https://rechtssammlung.sp.ethz.ch/Dokumente/510.21en.pdf} \\
\botrule
\end{tabularx}%
\end{sidewaystable}

\begin{sidewaystable}[]
\centering
\caption{The median harm experienced by papers indirectly citing retracted articles varies over time across different citation distances.}
\label{tab:tableS2}
\begin{tabularx}{0.95\columnwidth}{@{\extracolsep\fill}
w{c}{1.5cm}
w{c}{0.4cm}
ccccc
}
\toprule
Citation Distance &
  Nums &
  Y1 Harm (IQR) &
  Y2 Harm (IQR) &
  Y3 Harm (IQR) &
  Y4 Harm (IQR) &
  Y5 Harm (IQR) \\
  \midrule
2 &
  1469564 &
  \begin{tabular}[c]{@{}c@{}}23.16\\ (-48.67$\sim$72.32)\end{tabular} &
  \begin{tabular}[c]{@{}c@{}}21.22\\ (-47.63$\sim$67.02)\end{tabular} &
  \begin{tabular}[c]{@{}c@{}}24.18\\ (-44.96$\sim$69.16)\end{tabular} &
  \begin{tabular}[c]{@{}c@{}}27.38\\ (-42.30$\sim$71.87)\end{tabular} &
  \begin{tabular}[c]{@{}c@{}}31.19\\ (-39.37$\sim$75.21)\end{tabular} \\
3 &
  4972207 &
  \begin{tabular}[c]{@{}c@{}}32.07\\ (-38.57$\sim$90.13)\end{tabular} &
  \begin{tabular}[c]{@{}c@{}}29.16\\ (-37.85$\sim$74.78)\end{tabular} &
  \begin{tabular}[c]{@{}c@{}}31.57\\ (-35.64$\sim$76.29)\end{tabular} &
  \begin{tabular}[c]{@{}c@{}}34.68\\ (-33.29$\sim$79.31)\end{tabular} &
  \begin{tabular}[c]{@{}c@{}}37.99\\ (-30.51$\sim$82.67)\end{tabular} \\
4 &
  8319030 &
  \begin{tabular}[c]{@{}c@{}}39.36\\ (-33.93$\sim$100.00)\end{tabular} &
  \begin{tabular}[c]{@{}c@{}}34.96\\ (-33.51$\sim$85.54)\end{tabular} &
  \begin{tabular}[c]{@{}c@{}}37.02\\ (-31.43$\sim$86.38)\end{tabular} &
  \begin{tabular}[c]{@{}c@{}}40.00\\ (-29.26$\sim$100.00)\end{tabular} &
  \begin{tabular}[c]{@{}c@{}}42.79\\ (-26.59$\sim$100.00)\end{tabular} \\
5 &
  10729186 &
  \begin{tabular}[c]{@{}c@{}}44.42\\ (-31.37$\sim$100.00)\end{tabular} &
  \begin{tabular}[c]{@{}c@{}}38.80\\ (-31.26$\sim$100.00)\end{tabular} &
  \begin{tabular}[c]{@{}c@{}}40.71\\ (-29.19$\sim$100.00)\end{tabular} &
  \begin{tabular}[c]{@{}c@{}}43.55\\ (-26.99$\sim$100.00)\end{tabular} &
  \begin{tabular}[c]{@{}c@{}}46.67\\ (-24.27$\sim$100.00)\end{tabular} \\
\bottomrule
\end{tabularx}

\vspace{0.5cm}

\centering
\begin{tabularx}{0.95\columnwidth}{@{\extracolsep\fill}
w{c}{1.5cm}
w{c}{0.4cm}
ccccc
}
\toprule
 &
   &
  Y6 Harm (IQR) &
  Y7 Harm (IQR) &
  Y8 Harm (IQR) &
  Y8 Harm (IQR) &
  Y10 Harm (IQR) \\
  \midrule
 &
   &
  \begin{tabular}[c]{@{}c@{}}34.19\\ (-36.99$\sim$78.12)\end{tabular} &
  \begin{tabular}[c]{@{}c@{}}37.13\\ (-33.97$\sim$80.66)\end{tabular} &
  \begin{tabular}[c]{@{}c@{}}40.29\\ (-31.24$\sim$85.29)\end{tabular} &
  \begin{tabular}[c]{@{}c@{}}43.36\\ (-28.12$\sim$100.00)\end{tabular} &
  \begin{tabular}[c]{@{}c@{}}45.98\\ (-26.33$\sim$100.00)\end{tabular} \\
 &
   &
  \begin{tabular}[c]{@{}c@{}}40.71\\ (-28.53$\sim$88.21)\end{tabular} &
  \begin{tabular}[c]{@{}c@{}}43.61\\ (-25.80$\sim$100.00)\end{tabular} &
  \begin{tabular}[c]{@{}c@{}}46.76\\ (-23.06$\sim$100.00)\end{tabular} &
  \begin{tabular}[c]{@{}c@{}}49.48\\ (-20.48$\sim$100.00)\end{tabular} &
  \begin{tabular}[c]{@{}c@{}}52.21\\ (-18.37$\sim$100.00)\end{tabular} \\
 &
   &
  \begin{tabular}[c]{@{}c@{}}45.91\\ (-24.51$\sim$100.00)\end{tabular} &
  \begin{tabular}[c]{@{}c@{}}48.82\\ (-21.96$\sim$100.00)\end{tabular} &
  \begin{tabular}[c]{@{}c@{}}51.91\\ (-19.23$\sim$100.00)\end{tabular} &
  \begin{tabular}[c]{@{}c@{}}54.61\\ (-16.66$\sim$100.00)\end{tabular} &
  \begin{tabular}[c]{@{}c@{}}57.63\\ (-14.39$\sim$100.00)\end{tabular} \\
 &
   &
  \begin{tabular}[c]{@{}c@{}}49.95\\ (-22.05$\sim$100.00)\end{tabular} &
  \begin{tabular}[c]{@{}c@{}}53.04\\ (-19.53$\sim$100.00)\end{tabular} &
  \begin{tabular}[c]{@{}c@{}}56.04\\ (-16.73$\sim$100.00)\end{tabular} &
  \begin{tabular}[c]{@{}c@{}}59.25\\ (-13.88$\sim$100.00)\end{tabular} &
  \begin{tabular}[c]{@{}c@{}}62.84\\ (-11.50$\sim$100.00)\end{tabular} \\
\bottomrule
\end{tabularx}
\end{sidewaystable}

\begin{sidewaystable}[]
\centering
\caption{After removing duplicate statistics, the median harm experienced by papers indirectly citing retracted articles varies over time across different citation distances.}
\label{tab:tableS3}
\begin{tabularx}{0.95\columnwidth}{@{\extracolsep\fill}
w{c}{2cm}
w{c}{0.5cm}
ccccc
}
\toprule
Citation Distance &
  Nums &
  Y1 Harm (IQR) &
  Y2 Harm (IQR) &
  Y3 Harm (IQR) &
  Y4 Harm (IQR) &
  Y5 Harm (IQR) \\
  \midrule
2 &
  1374471 &
  \begin{tabular}[c]{@{}c@{}}24.22\\ (-46.76$\sim$73.04)\end{tabular} &
  \begin{tabular}[c]{@{}c@{}}22.19\\ (-45.93$\sim$67.56)\end{tabular} &
  \begin{tabular}[c]{@{}c@{}}25.07\\ (-43.40$\sim$69.66)\end{tabular} &
  \begin{tabular}[c]{@{}c@{}}28.23\\ (-40.58$\sim$72.42)\end{tabular} &
  \begin{tabular}[c]{@{}c@{}}31.99\\ (-37.70$\sim$75.69)\end{tabular} \\
3 &
  3529585 &
  \begin{tabular}[c]{@{}c@{}}36.21\\ (-34.05$\sim$100.00)\end{tabular} &
  \begin{tabular}[c]{@{}c@{}}32.87\\ (-33.52$\sim$78.68)\end{tabular} &
  \begin{tabular}[c]{@{}c@{}}35.04\\ (-31.53$\sim$79.92)\end{tabular} &
  \begin{tabular}[c]{@{}c@{}}37.95\\ (-29.31$\sim$82.99)\end{tabular} &
  \begin{tabular}[c]{@{}c@{}}40.80\\ (-26.68$\sim$89.29)\end{tabular} \\
4 &
  3342674 &
  \begin{tabular}[c]{@{}c@{}}52.11\\ (-25.65$\sim$100.00)\end{tabular} &
  \begin{tabular}[c]{@{}c@{}}44.62\\ (-26.37$\sim$100.00)\end{tabular} &
  \begin{tabular}[c]{@{}c@{}}45.93\\ (-24.65$\sim$100.00)\end{tabular} &
  \begin{tabular}[c]{@{}c@{}}48.51\\ (-22.46$\sim$100.00)\end{tabular} &
  \begin{tabular}[c]{@{}c@{}}51.52\\ (-20.23$\sim$100.00)\end{tabular} \\
5 &
  2397650 &
  \begin{tabular}[c]{@{}c@{}}73.48\\ (-20.36$\sim$100.00)\end{tabular} &
  \begin{tabular}[c]{@{}c@{}}55.39\\ (-21.84$\sim$100.00)\end{tabular} &
  \begin{tabular}[c]{@{}c@{}}56.38\\ (-20.00$\sim$100.00)\end{tabular} &
  \begin{tabular}[c]{@{}c@{}}59.59\\ (-17.35$\sim$100.00)\end{tabular} &
  \begin{tabular}[c]{@{}c@{}}63.74\\ (-15.02$\sim$100.00)\end{tabular} \\
\bottomrule
\end{tabularx}

\vspace{0.5cm}

\centering
\begin{tabularx}{0.95\columnwidth}{@{\extracolsep\fill}
w{c}{2cm}
w{c}{0.5cm}
ccccc
}
\toprule
 &
   &
  Y6 Harm (IQR) &
  Y7 Harm (IQR) &
  Y8 Harm (IQR) &
  Y8 Harm (IQR) &
  Y10 Harm (IQR) \\
  \midrule
 &
   &
  \begin{tabular}[c]{@{}c@{}}34.91\\ (-35.36$\sim$78.56)\end{tabular} &
  \begin{tabular}[c]{@{}c@{}}37.79\\ (-32.45$\sim$81.14)\end{tabular} &
  \begin{tabular}[c]{@{}c@{}}41.02\\ (-29.82$\sim$86.07)\end{tabular} &
  \begin{tabular}[c]{@{}c@{}}44.06\\ (-26.80$\sim$100.00)\end{tabular} &
  \begin{tabular}[c]{@{}c@{}}46.58\\ (-25.00$\sim$100.00)\end{tabular} \\
 &
   &
  \begin{tabular}[c]{@{}c@{}}43.72\\ (-24.81$\sim$100.00)\end{tabular} &
  \begin{tabular}[c]{@{}c@{}}46.69\\ (-22.09$\sim$100.00)\end{tabular} &
  \begin{tabular}[c]{@{}c@{}}49.69\\ (-19.34$\sim$100.00)\end{tabular} &
  \begin{tabular}[c]{@{}c@{}}52.29\\ (-17.04$\sim$100.00)\end{tabular} &
  \begin{tabular}[c]{@{}c@{}}55.16\\ (-14.86$\sim$100.00)\end{tabular} \\
 &
   &
  \begin{tabular}[c]{@{}c@{}}54.63\\ (-17.84$\sim$100.00)\end{tabular} &
  \begin{tabular}[c]{@{}c@{}}57.52\\ (-15.85$\sim$100.00)\end{tabular} &
  \begin{tabular}[c]{@{}c@{}}60.43\\ (-13.01$\sim$100.00)\end{tabular} &
  \begin{tabular}[c]{@{}c@{}}63.90\\ (-10.28$\sim$100.00)\end{tabular} &
  \begin{tabular}[c]{@{}c@{}}67.84\\ (-8.11$\sim$100.00)\end{tabular} \\
 &
   &
  \begin{tabular}[c]{@{}c@{}}68.53\\ (-11.89$\sim$100.00)\end{tabular} &
  \begin{tabular}[c]{@{}c@{}}74.10\\ (-9.00$\sim$100.00)\end{tabular} &
  \begin{tabular}[c]{@{}c@{}}82.61\\ (-5.72$\sim$100.00)\end{tabular} &
  \begin{tabular}[c]{@{}c@{}}100.00\\ (-2.52$\sim$100.00)\end{tabular} &
  \begin{tabular}[c]{@{}c@{}}100.00\\ (0.39$\sim$100.00)\end{tabular} \\
\bottomrule
\end{tabularx}

\end{sidewaystable}

\begin{sidewaystable}[]
\footnotesize
\centering
\caption{The median harm experienced by papers citing retracted articles, across different citation distances and published in journals with varying impact factors, varies over time after their publication (Part 1).}
\label{tab:tableS4}
\begin{tabularx}{0.95\columnwidth}{@{\extracolsep\fill}
w{c}{1.5cm}
w{c}{0.4cm}
w{c}{0.4cm}
ccccc
}
\toprule
Citation Distance &
  JIF &
  Nums &
  Y1 Harm (IQR) &
  Y2 Harm (IQR) &
  Y3 Harm (IQR) &
  Y4 Harm (IQR) &
  Y5 Harm (IQR) \\ \midrule
1 &
  0$\sim$3 &
  34814 &
  \begin{tabular}[c]{@{}c@{}}12.22\\ (-79.05$\sim$69.88)\end{tabular} &
  \begin{tabular}[c]{@{}c@{}}8.80\\ (-76.51$\sim$65.18)\end{tabular} &
  \begin{tabular}[c]{@{}c@{}}12.67\\ (-73.61$\sim$67.39)\end{tabular} &
  \begin{tabular}[c]{@{}c@{}}16.92\\ (-71.02$\sim$69.35)\end{tabular} &
  \begin{tabular}[c]{@{}c@{}}21.83\\ (-66.18$\sim$72.33)\end{tabular} \\
1 &
  3$\sim$5 &
  29091 &
  \begin{tabular}[c]{@{}c@{}}10.33\\ (-62.95$\sim$58.07)\end{tabular} &
  \begin{tabular}[c]{@{}c@{}}9.58\\ (-61.56$\sim$56.00)\end{tabular} &
  \begin{tabular}[c]{@{}c@{}}13.18\\ (-58.76$\sim$59.58)\end{tabular} &
  \begin{tabular}[c]{@{}c@{}}17.53\\ (-56.98$\sim$62.24)\end{tabular} &
  \begin{tabular}[c]{@{}c@{}}20.90\\ (-50.07$\sim$65.87)\end{tabular} \\
1 &
  5$\sim$10 &
  23000 &
  \begin{tabular}[c]{@{}c@{}}10.59\\ (-56.02$\sim$53.96)\end{tabular} &
  \begin{tabular}[c]{@{}c@{}}9.11\\ (-56.91$\sim$52.63)\end{tabular} &
  \begin{tabular}[c]{@{}c@{}}12.44\\ (-52.21$\sim$55.34)\end{tabular} &
  \begin{tabular}[c]{@{}c@{}}16.64\\ (-50.97$\sim$59.24)\end{tabular} &
  \begin{tabular}[c]{@{}c@{}}20.36\\ (-48.88$\sim$61.92)\end{tabular} \\
1 &
  10$\sim$20 &
  6273 &
  \begin{tabular}[c]{@{}c@{}}2.15\\ (-75.93$\sim$50.68)\end{tabular} &
  \begin{tabular}[c]{@{}c@{}}2.64\\ (-77.98$\sim$51.36)\end{tabular} &
  \begin{tabular}[c]{@{}c@{}}6.17\\ (-75.20$\sim$53.51)\end{tabular} &
  \begin{tabular}[c]{@{}c@{}}11.41\\ (-71.95$\sim$57.89)\end{tabular} &
  \begin{tabular}[c]{@{}c@{}}14.02\\ (-70.00$\sim$60.96)\end{tabular} \\
\bottomrule
\end{tabularx}

\vspace{0.3cm}

\centering
\begin{tabularx}{0.95\columnwidth}{@{\extracolsep\fill}
w{c}{1.5cm}
w{c}{0.4cm}
w{c}{0.4cm}
ccccc
}
\toprule
 &
   &
   &
  Y6 Harm (IQR) &
  Y7 Harm (IQR) &
  Y8 Harm (IQR) &
  Y9 Harm (IQR) &
  Y10 Harm (IQR) \\ \midrule
 &
   &
   &
  \begin{tabular}[c]{@{}c@{}}24.78\\ (-63.00$\sim$77.42)\end{tabular} &
  \begin{tabular}[c]{@{}c@{}}29.30\\ (-59.45$\sim$100.00)\end{tabular} &
  \begin{tabular}[c]{@{}c@{}}31.72\\ (-56.64$\sim$100.00)\end{tabular} &
  \begin{tabular}[c]{@{}c@{}}36.31\\ (-50.92$\sim$100.00)\end{tabular} &
  \begin{tabular}[c]{@{}c@{}}39.58\\ (-48.31$\sim$100.00)\end{tabular} \\
 &
   &
   &
  \begin{tabular}[c]{@{}c@{}}24.03\\ (-50.37$\sim$71.09)\end{tabular} &
  \begin{tabular}[c]{@{}c@{}}28.47\\ (-47.98$\sim$72.91)\end{tabular} &
  \begin{tabular}[c]{@{}c@{}}31.38\\ (-42.64$\sim$75.13)\end{tabular} &
  \begin{tabular}[c]{@{}c@{}}34.49\\ (-41.09$\sim$74.81)\end{tabular} &
  \begin{tabular}[c]{@{}c@{}}35.62\\ (-38.41$\sim$78.66)\end{tabular} \\
 &
   &
   &
  \begin{tabular}[c]{@{}c@{}}23.84\\ (-43.72$\sim$65.82)\end{tabular} &
  \begin{tabular}[c]{@{}c@{}}26.60\\ (-41.92$\sim$67.39)\end{tabular} &
  \begin{tabular}[c]{@{}c@{}}30.19\\ (-39.99$\sim$70.51)\end{tabular} &
  \begin{tabular}[c]{@{}c@{}}32.89\\ (-36.75$\sim$73.69)\end{tabular} &
  \begin{tabular}[c]{@{}c@{}}35.20\\ (-34.05$\sim$75.71)\end{tabular} \\
 &
   &
   &
  \begin{tabular}[c]{@{}c@{}}18.53\\ (-62.80$\sim$64.32)\end{tabular} &
  \begin{tabular}[c]{@{}c@{}}22.31\\ (-58.95$\sim$67.40)\end{tabular} &
  \begin{tabular}[c]{@{}c@{}}25.61\\ (-56.99$\sim$69.09)\end{tabular} &
  \begin{tabular}[c]{@{}c@{}}27.37\\ (-52.30$\sim$72.05)\end{tabular} &
  \begin{tabular}[c]{@{}c@{}}32.63\\ (-49.45$\sim$73.31)\end{tabular} \\
\bottomrule
\end{tabularx}
\end{sidewaystable}

\addtocounter{table}{-1}

\begin{sidewaystable}[]
\footnotesize
\centering
\caption{The median harm experienced by papers citing retracted articles, across different citation distances and published in journals with varying impact factors, varies over time after their publication (Part 2).}
\label{tab:tableS4c}
\begin{tabularx}{0.95\columnwidth}{@{\extracolsep\fill}
w{c}{1.5cm}
w{c}{0.4cm}
w{c}{0.4cm}
ccccc
}
\toprule
Citation Distance &
  JIF &
  Nums &
  Y1 Harm (IQR) &
  Y2 Harm (IQR) &
  Y3 Harm (IQR) &
  Y4 Harm (IQR) &
  Y5 Harm (IQR) \\ \midrule
4 &
  0$\sim$3 &
  3056591 &
  \begin{tabular}[c]{@{}c@{}}37.88\\ (-37.48$\sim$100.00)\end{tabular} &
  \begin{tabular}[c]{@{}c@{}}33.61\\ (-36.16$\sim$100.00)\end{tabular} &
  \begin{tabular}[c]{@{}c@{}}35.28\\ (-34.15$\sim$86.87)\end{tabular} &
  \begin{tabular}[c]{@{}c@{}}37.56\\ (-32.02$\sim$100.00)\end{tabular} &
  \begin{tabular}[c]{@{}c@{}}40.27\\ (-29.76$\sim$100.00)\end{tabular} \\
4 &
  3$\sim$5 &
  1533994 &
  \begin{tabular}[c]{@{}c@{}}31.18\\ (-28.62$\sim$72.95)\end{tabular} &
  \begin{tabular}[c]{@{}c@{}}30.43\\ (-27.10$\sim$70.49)\end{tabular} &
  \begin{tabular}[c]{@{}c@{}}32.77\\ (-25.35$\sim$72.98)\end{tabular} &
  \begin{tabular}[c]{@{}c@{}}35.66\\ (-23.32$\sim$75.29)\end{tabular} &
  \begin{tabular}[c]{@{}c@{}}38.61\\ (-21.54$\sim$76.80)\end{tabular} \\
4 &
  5$\sim$10 &
  887475 &
  \begin{tabular}[c]{@{}c@{}}30.11\\ (-26.91$\sim$67.85)\end{tabular} &
  \begin{tabular}[c]{@{}c@{}}29.44\\ (-25.26$\sim$67.68)\end{tabular} &
  \begin{tabular}[c]{@{}c@{}}32.27\\ (-23.35$\sim$69.84)\end{tabular} &
  \begin{tabular}[c]{@{}c@{}}35.18\\ (-21.54$\sim$71.34)\end{tabular} &
  \begin{tabular}[c]{@{}c@{}}37.65\\ (-19.49$\sim$74.04)\end{tabular} \\
4 &
  10$\sim$20 &
  236740 &
  \begin{tabular}[c]{@{}c@{}}29.57\\ (-34.41$\sim$71.73)\end{tabular} &
  \begin{tabular}[c]{@{}c@{}}30.67\\ (-32.64$\sim$72.32)\end{tabular} &
  \begin{tabular}[c]{@{}c@{}}33.98\\ (-30.13$\sim$75.17)\end{tabular} &
  \begin{tabular}[c]{@{}c@{}}36.95\\ (-28.01$\sim$77.26)\end{tabular} &
  \begin{tabular}[c]{@{}c@{}}39.68\\ (-25.20$\sim$79.90)\end{tabular} \\
\bottomrule
\end{tabularx}

\vspace{0.3cm}

\centering
\begin{tabularx}{0.95\columnwidth}{@{\extracolsep\fill}
w{c}{1.5cm}
w{c}{0.4cm}
w{c}{0.4cm}
ccccc
}
\toprule
 &
   &
   &
  Y6 Harm (IQR) &
  Y7 Harm (IQR) &
  Y8 Harm (IQR) &
  Y9 Harm (IQR) &
  Y10 Harm (IQR) \\ \midrule
 &
   &
   &
  \begin{tabular}[c]{@{}c@{}}43.04\\ (-27.99$\sim$100.00)\end{tabular} &
  \begin{tabular}[c]{@{}c@{}}45.77\\ (-25.44$\sim$100.00)\end{tabular} &
  \begin{tabular}[c]{@{}c@{}}48.65\\ (-22.77$\sim$100.00)\end{tabular} &
  \begin{tabular}[c]{@{}c@{}}50.98\\ (-20.46$\sim$100.00)\end{tabular} &
  \begin{tabular}[c]{@{}c@{}}53.17\\ (-18.68$\sim$100.00)\end{tabular} \\
 &
   &
   &
  \begin{tabular}[c]{@{}c@{}}40.68\\ (-19.95$\sim$77.92)\end{tabular} &
  \begin{tabular}[c]{@{}c@{}}42.90\\ (-18.06$\sim$78.91)\end{tabular} &
  \begin{tabular}[c]{@{}c@{}}46.06\\ (-15.78$\sim$80.73)\end{tabular} &
  \begin{tabular}[c]{@{}c@{}}48.42\\ (-13.97$\sim$89.25)\end{tabular} &
  \begin{tabular}[c]{@{}c@{}}50.99\\ (-11.99$\sim$100.00)\end{tabular} \\
 &
   &
   &
  \begin{tabular}[c]{@{}c@{}}39.79\\ (-17.82$\sim$76.95)\end{tabular} &
  \begin{tabular}[c]{@{}c@{}}42.46\\ (-15.82$\sim$79.21)\end{tabular} &
  \begin{tabular}[c]{@{}c@{}}44.96\\ (-13.84$\sim$80.50)\end{tabular} &
  \begin{tabular}[c]{@{}c@{}}46.84\\ (-12.08$\sim$81.69)\end{tabular} &
  \begin{tabular}[c]{@{}c@{}}48.91\\ (-10.62$\sim$82.96)\end{tabular} \\
 &
   &
   &
  \begin{tabular}[c]{@{}c@{}}42.51\\ (-23.08$\sim$81.81)\end{tabular} &
  \begin{tabular}[c]{@{}c@{}}44.77\\ (-20.68$\sim$83.14)\end{tabular} &
  \begin{tabular}[c]{@{}c@{}}47.03\\ (-18.20$\sim$84.87)\end{tabular} &
  \begin{tabular}[c]{@{}c@{}}49.43\\ (-15.69$\sim$86.67)\end{tabular} &
  \begin{tabular}[c]{@{}c@{}}51.11\\ (-14.03$\sim$87.48)\end{tabular} \\
\bottomrule
\end{tabularx}
\end{sidewaystable}

\begin{sidewaystable}[]
\centering
\caption{Before and after retraction, the median harm experienced by papers citing retracted articles varies over time across different citation distances.}
\label{tab:tableS5}
\begin{tabularx}{0.95\columnwidth}{@{\extracolsep\fill}
w{c}{2cm}
w{c}{2cm}
w{c}{0.4cm}
ccccc
}
\toprule
Citation Distance &
  Citation Timing &
  Nums &
  Y1 Harm (IQR) &
  Y2 Harm (IQR) &
  Y3 Harm (IQR) &
  Y4 Harm (IQR) &
  Y5 Harm (IQR) \\ \midrule
\multirow{2}{*}{\centering 1} &
  After Retraction &
  27155 &
  \begin{tabular}[c]{@{}c@{}}18.40\\ (-65.16$\sim$74.86)\end{tabular} &
  \begin{tabular}[c]{@{}c@{}}15.39\\ (-65.83$\sim$68.14)\end{tabular} &
  \begin{tabular}[c]{@{}c@{}}18.38\\ (-63.45$\sim$70.07)\end{tabular} &
  \begin{tabular}[c]{@{}c@{}}22.28\\ (-59.67$\sim$72.98)\end{tabular} &
  \begin{tabular}[c]{@{}c@{}}27.18\\ (-53.77$\sim$77.30)\end{tabular} \\
 &
  Before Retraction &
  83405 &
  \begin{tabular}[c]{@{}c@{}}11.36\\ (-68.77$\sim$64.20)\end{tabular} &
  \begin{tabular}[c]{@{}c@{}}8.93\\ (-69.30$\sim$59.84)\end{tabular} &
  \begin{tabular}[c]{@{}c@{}}12.28\\ (-65.84$\sim$62.49)\end{tabular} &
  \begin{tabular}[c]{@{}c@{}}16.67\\ (-63.94$\sim$65.39)\end{tabular} &
  \begin{tabular}[c]{@{}c@{}}20.89\\ (-58.28$\sim$69.18)\end{tabular} \\
\multirow{2}{*}{\centering 2} &
  After Retraction &
  741602 &
  \begin{tabular}[c]{@{}c@{}}21.43\\ (-50.37$\sim$69.91)\end{tabular} &
  \begin{tabular}[c]{@{}c@{}}19.28\\ (-50.09$\sim$64.79)\end{tabular} &
  \begin{tabular}[c]{@{}c@{}}22.40\\ (-47.42$\sim$67.33)\end{tabular} &
  \begin{tabular}[c]{@{}c@{}}25.58\\ (-44.77$\sim$70.19)\end{tabular} &
  \begin{tabular}[c]{@{}c@{}}29.56\\ (-41.47$\sim$73.38)\end{tabular} \\
 &
  Before Retraction &
  555586 &
  \begin{tabular}[c]{@{}c@{}}24.51\\ (-44.96$\sim$72.50)\end{tabular} &
  \begin{tabular}[c]{@{}c@{}}22.36\\ (-44.85$\sim$67.58)\end{tabular} &
  \begin{tabular}[c]{@{}c@{}}25.14\\ (-42.18$\sim$69.49)\end{tabular} &
  \begin{tabular}[c]{@{}c@{}}28.12\\ (-39.55$\sim$71.91)\end{tabular} &
  \begin{tabular}[c]{@{}c@{}}31.71\\ (-37.09$\sim$75.15)\end{tabular} \\
\bottomrule
\end{tabularx}

\vspace{0.3cm}

\centering
\begin{tabularx}{0.95\columnwidth}{@{\extracolsep\fill}
w{c}{2cm}
w{c}{2cm}
w{c}{0.4cm}
ccccc
}
\toprule
 &
   &
   &
  Y6 Harm (IQR) &
  Y7 Harm (IQR) &
  Y8 Harm (IQR) &
  Y9 Harm (IQR) &
  Y10 Harm (IQR) \\ \midrule
 &
   &
   &
  \begin{tabular}[c]{@{}c@{}}31.09\\ (-52.07$\sim$80.44)\end{tabular} &
  \begin{tabular}[c]{@{}c@{}}35.27\\ (-47.95$\sim$85.11)\end{tabular} &
  \begin{tabular}[c]{@{}c@{}}37.65\\ (-44.51$\sim$100.00)\end{tabular} &
  \begin{tabular}[c]{@{}c@{}}42.50\\ (-39.39$\sim$100.00)\end{tabular} &
  \begin{tabular}[c]{@{}c@{}}45.14\\ (-36.07$\sim$100.00)\end{tabular} \\
 &
   &
   &
  \begin{tabular}[c]{@{}c@{}}24.24\\ (-56.17$\sim$72.41)\end{tabular} &
  \begin{tabular}[c]{@{}c@{}}27.81\\ (-53.10$\sim$75.46)\end{tabular} &
  \begin{tabular}[c]{@{}c@{}}31.34\\ (-49.88$\sim$78.10)\end{tabular} &
  \begin{tabular}[c]{@{}c@{}}34.39\\ (-45.98$\sim$81.83)\end{tabular} &
  \begin{tabular}[c]{@{}c@{}}37.53\\ (-43.40$\sim$98.78)\end{tabular} \\
 &
   &
   &
  \begin{tabular}[c]{@{}c@{}}32.95\\ (-38.91$\sim$76.63)\end{tabular} &
  \begin{tabular}[c]{@{}c@{}}35.68\\ (-36.15$\sim$78.80)\end{tabular} &
  \begin{tabular}[c]{@{}c@{}}38.67\\ (-33.20$\sim$82.18)\end{tabular} &
  \begin{tabular}[c]{@{}c@{}}42.28\\ (-29.95$\sim$93.02)\end{tabular} &
  \begin{tabular}[c]{@{}c@{}}44.52\\ (-28.40$\sim$100.00)\end{tabular} \\
 &
   &
   &
  \begin{tabular}[c]{@{}c@{}}34.58\\ (-34.90$\sim$77.98)\end{tabular} &
  \begin{tabular}[c]{@{}c@{}}37.41\\ (-32.02$\sim$80.44)\end{tabular} &
  \begin{tabular}[c]{@{}c@{}}40.68\\ (-29.48$\sim$84.34)\end{tabular} &
  \begin{tabular}[c]{@{}c@{}}43.46\\ (-26.80$\sim$100.00)\end{tabular} &
  \begin{tabular}[c]{@{}c@{}}46.13\\ (-24.60$\sim$100.00)\end{tabular} \\
\bottomrule
\end{tabularx}

\end{sidewaystable}

\end{document}